\documentclass[draftclsnofoot,onecolumn]{IEEEtran}
\usepackage{cite,amssymb,graphicx,psfrag,upgreek}
\usepackage[cmex10]{amsmath}
\usepackage[tight,footnotesize]{subfigure}
\usepackage{MyCommands}

\begin{document}

\title{Optimal model-based beamforming and independent steering for spherical loudspeaker arrays}

\author{Boaz Rafaely and Dima Khaykin
\\ Department of Electrical and Computer Engineering, Ben-Gurion University of the Negev \\ Beer-Sheva 84105, Israel \{br,khaykin\}@ee.bgu.ac.il, 6 February 2011
}

\maketitle

\footnote{Copyright (c) 2010 IEEE. Personal use of this material is permitted. However, permission to use this material for any other purposes must be obtained from the IEEE by sending a request to pubs-permissions@ieee.org.}

\begin{abstract}
Spherical loudspeaker arrays have been recently studied for directional sound radiation, where the compact arrangement of the loudspeaker units around a sphere facilitated the control of sound radiation in three-dimensional space. Directivity of sound radiation, or beamforming, was achieved by driving each loudspeaker unit independently, where the design of beamforming weights was typically achieved by numerical optimization with reference to a given desired beam pattern. This is in contrast to the methods already developed for microphone arrays in general and spherical microphone arrays in particular, where beamformer weights are designed to satisfy a wider range of objectives, related to directivity, robustness, and side-lobe level, for example. This paper presents the development of a physical-model-based, optimal beamforming framework for spherical loudspeaker arrays, similar to the framework already developed for spherical microphone arrays, facilitating efficient beamforming in the spherical harmonics domain, with independent steering. In particular, it is shown that from a beamforming perspective, the spherical loudspeaker array is similar to the spherical microphone array with microphones arranged around a rigid sphere. Experimental investigation validates the theoretical framework of beamformer design.
\end{abstract}

\section{Introduction}

Spherical loudspeaker arrays, composed of a set of loudspeaker units mounted on the surface of a sphere, operating as a multiple-channel sound source, have been recently studied for applications such as electro-acoustic music performance, synthesizing the radiation pattern of musical instruments \cite{warusfel2001,avizienis2006}, and active control of sound \cite{rafaely2009jasa}.
A physical model of the loudspeaker array has been developed \cite{zotter2007ICA,rafaely2009jasa}, as a rigid sphere with vibrating caps mounted on its surface, employing spherical harmonics to describe caps vibration and sound radiation \cite{williams1999}. At low frequencies, array directivity can be represented as a linear combination of spherical harmonics basis functions \cite{pasqual2010}, and by additionally including models of the loudspeaker units \cite{pollow2009}, array weights can be designed to achieve a desired direcitivity function. A comprehensive review of previous work concerning spherical loudspeaker arrays has been presented recently \cite{pasqual2010}.

Although useful in generating spherical harmonics based beam patterns, the methods presented in previous work posses the following shortcomings:

\begin{enumerate}
  \item Typically, beam-pattern matching was the sole design objective, and so robustness against noise and uncertainty was not introduced, which can degrade performance in practical systems.
  \item No simple steering of the beam pattern was presented, and so re-calculation of the beam pattern is typically required to realize steering.
  \item Alternative design approaches to the one previously presented, i.e. numerical fitting to a desired directivity function, may be of interest. These include multiple-objective designs; optimal designs; and analytical designs that produce closed-form expressions for the beamforming weights. However, a framework to apply these designs to spherical loudspeaker array was not presented.
\end{enumerate}

This paper presents a beamforming design framework for spherical loudspeaker arrays that overcomes the shortcoming presented above. The design framework is based on a physical model of the spherical loudspeaker array, presented in section \ref{SEC:model}. In this model the spherical loudspeaker array is represented by a rigid sphere with a set of caps mounted on its surface, representing the vibration of the diaphragm of the loudspeaker units, which is then further simplified to a spherical source with radial velocity represented in the spherical harmonics domain. This model is then used to develop the fundamental beamforming equations in section \ref{SEC:beamforming}, both in the space domain by weighting caps velocities, and more generally in the spherical harmonics domain. Section \ref{SEC:farfield} presents the beamforming formulation for far-field, axis-symmetric radiation, which is central to this paper. It is shown that the resulting beamforming problem is almost identical to the beamforming problem of a spherical microphone array with microphones arranged around a rigid sphere. The latter has been recently introduced \cite{meyer2002}, has been studied extensively since, with well investigated analysis of performance \cite{rafaely2005tsap}, and with a range of beamforming methods developed \cite{rafaely2010book}. The novel result of the similarity between the two arrays leads directly to the development of a beamforming design method for spherical loudspeaker arrays that is based on the framework developed for spherical microphone arrays. A formulation of measures for array  directivity index and robustness are presented in section \ref{SEC:directivity}, after which optimal beamformers with simple steering for the spherical loudspeaker array are developed in section \ref{SEC:optimal}, including maximum directivity, maximum robustness, and Dolph-Chebyshev, as examples. Experimental investigation of beamforming with a real array having 12 loudspeaker units, measured in an anechoic chamber, concludes the paper.

\section{Sound radiation from spherical sources}\label{SEC:model}

Sound radiation from spherical sources is reviewed in this section. A spherical source is modeled as a rigid sphere of radius $r_0$ with $L$ spherical caps, representing loudspeaker units, positioned on its surface at locations $(\theta_l,\phi_l)$, each imposing a constant radial surface velocity of $v_l, \ l=1,...,L$, at the surface segment they cover \cite{rafaely2009jasa,zotter2007DAGA}. Here $\theta_l$ represent elevation angle, measured down from the z-axis, and $\phi_l$ represent azimuth angle, measured on the x-y plane away from the x-axis towards the y-axis, defining a spherical coordinate system \cite{arfken2001}. The radial velocity of the sphere surface at wave number $k$, $u(k,r_0,\theta,\phi)$, is composed of contributions from all $L$ caps. The spherical Fourier transform of the radial velocity, $u_{nm}(k,r_0)$, is defined as \cite{arfken2001}:

\begin{equation}\label{EQ:unmSFT}
    u_{nm}(k,r_0)=\int_0^{2\pi}\int_0^\pi u(k,r_0,\theta,\phi) [Y_n^m(\theta,\phi)]^* \sin\theta d\theta d\phi
\end{equation}

with $Y_n^m(\cdot,\cdot)$ the spherical harmonics of order $n$ and degree $m$. After deriving the spherical Fourier transform of the radial velocity due to a single cap and adding the contributions from all $L$ caps, Eq. (\ref{EQ:unmSFT}) reduces to \cite{rafaely2009jasa}:

\begin{equation}\label{EQ:unmCAPS}
u_{nm}(k,r_0) = g_n \sum_{l=1}^{L} v_l(k) [Y_n^m(\theta_l,\phi_l)]^*
\end{equation}

with
\begin{equation}\label{EQ:gn}
g_n \equiv \frac{4\pi^2}{2n+1}\left[P_{n-1}(\cos\alpha)-P_{n+1}(\cos\alpha)\right],
\end{equation}

and with $P_n(\cdot)$ the Legendre polynomial, and $\alpha$ the aperture angle of each spherical cap.

Given the radial velocity over the sphere surface, the sound pressure $p(k,r,\theta,\phi)$ away from the source, is computed by \cite{williams1999}:

\begin{equation}\label{EQ:pressure-unm}
p(k,r,\theta,\phi) = i\rho_0 c \sum_{n=0}^{\infty}\sum_{m=-n}^{n}\frac{h_n(kr)}{h_n^{'}(kr_0)} u_{nm}(k,r_0)  Y_n^m(\theta,\phi),
\end{equation}

with $c$ the speed of sound, $\rho_0$ air density, $i=\sqrt{-1}$, and $h_n(\cdot)$ and $h'_n(\cdot)$ are the spherical Hankel function of the first kind of order $n$, and it's derivative, respectively \cite{arfken2001}. Now, the spherical Fourier transform of the sound pressure, $p_{nm}(k,r)$, can be written as:

\begin{equation}\label{EQ:pressureSH}
p_{nm}(k,r) = i\rho_0 c \frac{h_n(kr)}{h'_n(kr_0)} u_{nm}(k,r_0)
\end{equation}

Equations (\ref{EQ:pressure-unm}), (\ref{EQ:unmCAPS}) and (\ref{EQ:gn}) can now be used to represent the sound pressure radiated by the spherical source, given the velocity of each spherical cap, or loudspeaker unit. It is worth noting that with $L$ spherical caps constructing the source, only $L$ spherical harmonics in $u_{nm}$ and $p_{nm}$ can be independently controlled, typically taking the first $(N+1)^2$ harmonics, such that

\begin{equation}\label{EQ:NL}
(N+1)^2\leq L
\end{equation}

with $n\leq N$, and $-n\leq m \leq n$. Also note that by controlling the radial velocity of $L$ caps, and assuming control over spherical harmonics of orders $n\le N$, the higher order harmonics $N+1$ and above cannot be controlled. However, sufficiently away from the source, at distances that satisfy $kr>>N$, the higher order harmonics are significantly attenuated by the term $h_n(kr)/h'_n(kr_0)$, and can be neglected \cite{rafaely2009jasa}. This means that although in practice source control is achieved through control over caps velocity, one can also assume a direct control over $u_{nm}$ at orders $n\le N$, with good accuracy.

\section{beamforming with a spherical source} \label{SEC:beamforming}

Beamforming with spherical sources is employed with the aim of controlling the directivity pattern of the sound radiated from the source. This is achieved by weighting the source signal $s(k)$ with weights $w_l(k)$ before driving the caps velocity, or loudspeaker units in practice, such that,

\begin{equation}\label{EQ:vl}
v_l(k) = w_l(k) s(k),\, l=1,...,L
\end{equation}

Now, repeating the derivation in Eqs. (\ref{EQ:unmSFT}), (\ref{EQ:unmCAPS}) and (\ref{EQ:gn}), but this time with $w_l(k)$, $w_{nm}(k)$ and $w(k,\theta,\phi)$ replacing $v_l(k)$, $u_{nm}(k,r_0)$ and $u(k,r_0,\theta,\phi)$, respectively, and using Eq. (\ref{EQ:vl}), the following holds:

\begin{equation}\label{EQ:wnm}
w_{nm}(k) = g_n \sum_{l=1}^{L} w_l(k) [Y_n^m(\theta_l,\phi_l)]^*
\end{equation}

and

\begin{equation}\label{EQ:unm-wnm}
u_{nm}(k,r_0) = s(k) w_{nm}(k)
\end{equation}

with $w(k,\theta,\phi)$ representing the beamforming weight function as a continuous function over the sphere surface. Following the same argument as presented in section \ref{SEC:model}, one can assume control over $w_{nm}(k)$, although in practice beamforming is achieved through a direct control over $w_l(k)$.

Now, the pressure away from the source can be written in terms of the beamforming weights by substituting Eq. (\ref{EQ:unm-wnm}) in Eq. (\ref{EQ:pressure-unm}),

\begin{equation}\label{EQ:pressure-wnm}
p(k,r,\theta,\phi) = i\rho_0 c s(k) \sum_{n=0}^{\infty}\sum_{m=-n}^{n}\frac{h_n(kr)}{h_n^{'}(kr_0)} w_{nm}(k,r_0)  Y_n^m(\theta,\phi).
\end{equation}

Following Eqs. (\ref{EQ:pressure-wnm}) and (\ref{EQ:wnm}), beamforming design requires the computation of weights, $w_{nm}(k)$ or $w_l(k)$, such that the radiated sound pressure maintains some given design criterion. These equations, or measured versions of them, have been previously employed in a numerical design framework, for computing beamforming weights for spherical sources. The next section presents some further derivations, that will facilitate an analytical, rather than numerical design of beamformers for spherical sources, in a manner similar to spherical microphone arrays.

\section{Axis-symmetric far-field beamforming}\label{SEC:farfield}

An efficient formulation for far-field beamforming is derived in this section by constraining the radiated far-field sound pressure to be rotationally symmetric around the look direction, similar to the approach taken for spherical microphone array beamforming \cite{meyer2002}. We first assume that a far-field beam pattern is required, which is the case in most applications involving sound radiated into large rooms, such as music halls and video conferencing rooms. Far-field in the context of this work means that $kr>>N$, where $N$ is the highest order controlled by the source. In this case the following large-argument approximation can be employed \cite{williams1999}:

\begin{equation}\label{EQ:largekr}
h_n(kr) \approx (-i)^{n+1} \frac{e^{ikr}}{kr}
\end{equation}

Also, we introduce the Wronskian relation \cite{williams1999}:

\begin{equation}\label{EQ:wronskian}
j_n(kr)h'_n(kr) - j'_n(kr) h_n(kr) = \frac{i}{(kr)^{2}}
\end{equation}

which is rearranged as follows:

\begin{equation}\label{EQ:wronskian2}
\frac{1}{h'_n(kr)} = -i(kr)^2 \left[j_n(kr) - \frac{j'_n(kr)}{h'_n(kr)} h_n(kr)\right]
\end{equation}

We further denote for notation simplicity:

\begin{equation}\label{EQ:bn}
b_n(kr) \equiv i\rho_0 c k r^2 (-i)^n \left[j_n(kr) - \frac{j'_n(kr)}{h'_n(kr)} h_n(kr)\right]
\end{equation}

Substituting Eqs. (\ref{EQ:largekr}), (\ref{EQ:wronskian2}) and (\ref{EQ:bn}) into Eq. (\ref{EQ:pressure-wnm}), the far-field sound pressure can be written as:

\begin{equation}\label{EQ:pressure-wnmFF}
p(k,r,\theta,\phi) = \frac{e^{ikr}}{r} s(k)  \sum_{n=0}^{\infty}\sum_{m=-n}^{n} b_n(kr_0) w_{nm}(k,r_0)  Y_n^m(\theta,\phi).
\end{equation}

In the next step of this derivation, we remove the dependance on $r$ by considering the directivity function, or beam pattern $B$, computed by normalizing the far-field sound pressure with a factor of $re^{-ikr}$ \cite{williams1999}, and assuming a unit input signal $s(k)=1$,

\begin{equation}\label{EQ:B-wnmFF}
B(k,\theta,\phi) = \sum_{n=0}^{\infty}\sum_{m=-n}^{n} b_n(kr_0) w_{nm}(k,r_0)  Y_n^m(\theta,\phi).
\end{equation}

We now make a further simplification by considering axis-symmetric beam patterns, in a way similar to spherical microphone array beamforming \cite{rafaely2005tsap}, by selecting weights as follows:

\begin{equation}\label{EQ:wnm-dn}
w_{nm}(k) = \frac{d_n(k)}{b_n(kr_0)} [Y_n^m(\theta_0,\phi_0)]^*
\end{equation}

where $d_n(k)$ is the one-dimensional axis-symmetric beamforming weighting function, and $(\theta_0,\phi_0)$ is the look direction, forming the axis of symmetry. By substituting Eq. (\ref{EQ:wnm-dn}) in Eq. (\ref{EQ:B-wnmFF}), and using the spherical harmonics addition theorem  \cite{arfken2001}, the far-field directivity function can be rewritten as:

\begin{equation}\label{EQ:B-dnFF}
B(k,\Theta) = \sum_{n=0}^{N} d_n(k) \frac{2n+1}{4\pi} P_n(\cos\Theta)
\end{equation}

where $\Theta$ is the angle between the look direction $(\theta_0,\phi_0)$ and the direction of radiated sound, $(\theta,\phi)$, defined as:

\begin{equation}\label{EQ:Theta}
\cos\Theta = \cos\theta_0 \cos\theta + \cos(\phi_0-\phi) \sin\theta_0 \sin\theta.
\end{equation}

Several interesting observations can be made regarding the derivation in this section:

\begin{itemize}
  \item Eq. (\ref{EQ:B-dnFF}) representing the beam pattern for the spherical source is exactly the same as the beam pattern equation for spherical microphone arrays \cite{rafaely2005tsap}. A wide range of analytical beam pattern design methods have been developed for the latter, and will be proposed in this paper for beamforming with the spherical source.
  \item The term $b_n$ in Eq. (\ref{EQ:bn}) is very similar to the same term derived for spherical microphone arrays designed around a rigid sphere. In both cases, $b_n$ represents the dynamics of sound propagation around a rigid sphere, such that a division by $b_n$ turns the beam pattern independent on the spherical array configuration. It is therefore expected that beamforming with a spherical loudspeaker array as formulated in this paper will poses a similar behavior to beamforming with a spherical microphone array with a rigid sphere configuration.
  \item The weights $w_l$ are assumed to control caps velocity, $v_l$. In practice, the weights will control the signal driving the loudspeaker units, i.e. voltage input rather than velocity input. For typical moving-coil loudspeakers, operating above the mechanical cut-off frequency and below the radiation cut-off, the voltage is proportional to the frequency times the velocity \cite{kinsler1999}, such that in practice $k s(k)$ can be considered as directly proportional to the voltage signal if $s(k)$ is the velocity signal, and so the dependance on $k$ in Eqs. (\ref{EQ:bn}), (\ref{EQ:pressure-wnmFF}) and (\ref{EQ:B-wnmFF}) is removed, making the system models for the spherical loudspeaker and microphone arrays even more similar.
\end{itemize}

\section{Directivity index and robustness}\label{SEC:directivity}

Beamformer design typically involves achieving a desired directivity, while maintaining necessary robustness constraints \cite{vantrees2002}. A common measure for array performance is the directivity factor, calculated as the directivity at the look direction, relative to the directional average of the directivity function \cite{vantrees2002}:

\begin{equation} \label{EQ:Qintegral}
Q = \frac{\abs{B(k,\theta_0,\phi_0)}^2}{\frac{1}{4\pi} \int_{0}^{2\pi} \int_{0}^{\pi}  \abs{B(k,\theta,\phi)}^2 \sin\theta d\theta d\phi}
\end{equation}

The directivity factor for the spherical source can be derived by substituting Eq. (\ref{EQ:B-dnFF}) into Eq. (\ref{EQ:Qintegral}). Note that Eq. (\ref{EQ:B-dnFF}) is identical to the directivity function of the spherical microphone array \cite{rafaely2005spl}. The directivity factor, as derived in \cite{rafaely2005spl}, is therefore:

\begin{equation}\label{EQ:Qsum}
Q = \frac{\abs{\sum_{n=0}^N d_{n}(k)(2n+1)}^2}{\sum_{n=0}^N \abs{d_n(k)}^2 (2n+1)}
\end{equation}

The Directivity Index (DI) is now defined as $10\log_{10}Q$.

Another important measure is array robustness, which is a measure of the system sensitivity to noise, errors, uncertainties and perturbations. A common measure of robustness relates to the inverse of the 2-norm of the array weights, assuming the direcitvity function in the look direction is unity. The latter constraint is referred to as distortionless response. This measure is exactly the white-noise gain for sensor arrays, but is also considered as a general measure for robustness \cite{vantrees2002}. We adopt the same measure for the spherical source. We use the term white-noise gain (WNG) although in the context of this work it refers to general robustness. The WNG can be calculated by normalizing the 2-norm of the coefficients $w_{nm}$ to satisfy the distortionless response constraint with reference to Eq. (\ref{EQ:B-wnmFF}):

\begin{equation}\label{EQ:WNG-wnm}
\mathrm{WNG} = \frac{\left|\sum_{n=0}^{\infty}\sum_{m=-n}^{n} b_n(kr_0) w_{nm}(k,r_0)  Y_n^m(\theta_0,\phi_0)\right|^2}{\sum_{n=0}^N \sum_{m=-n}^n |w_{nm}(k)|^2}
\end{equation}

Substituting Eq. (\ref{EQ:wnm-dn}), and using the spherical harmonics addition theorem, we get:

\begin{equation}\label{EQ:WNG-dn}
\mathrm{WNG} = \frac{\left|\sum_{n=0}^{\infty} d_n(k)  (2n+1) \right|^2}{\sum_{n=0}^N \frac{|d_n(k)|^2}{|b_n(kr_0)|^2} (2n+1)}
\end{equation}

This result is equivalent to the WNG calculated for spherical microphone arrays \cite{rafaely2005spl}, although the function involved, e.g. $b_n(kr_0)$ are only equivalent up to some frequency-dependant constant, as discussed above.

\section{Optimal beamforming}\label{SEC:optimal}

Having developed expressions for the spherical source concerning directivity and WNG, in this section some optimal beamforming methods are proposed, which have analytical, or closed-form solutions, as opposed to most current methods for spherical source beamforming that use numerical optimization.

\subsection{Maximum Directivity}\label{SEC:maxQ}

This beamforming method aims to find the beamforming weights $d_n$ that maximize the directivity factor of a given spherical loudspeaker array, or spherical source. First, the problem is formulated in a matrix form and then the weights are derived that maximize the directivity factor.

The beamforming weights vector at wave number $k$ is defined as:

\begin{equation}\label{EQ:dnMAT}
\mathbf{d} = \left[d_0(k), d_1(k),..., d_N(k) \right]^T
\end{equation}

The following $(N+1)\times 1$ vector of coefficients, with the $n$-th element given by $2n+1$, is also defined:

\begin{equation}\label{EQ:2n}
\mathbf{a} = \left[1, 3,..., 2N+1 \right]^T
\end{equation}

such that the directivity factor, Eq. (\ref{EQ:Qsum}) can be written in a matrix form as:

\begin{equation}\label{EQ:QRayleigh}
Q = \frac{\mathbf{d}^H [\mathbf{a}^T \mathbf{a}] \mathbf{d}}{\mathbf{d}^H [\mathrm{diag}(\mathbf{a})] \mathbf{d}}
\end{equation}

Equation (\ref{EQ:QRayleigh}) represents a generalized Rayleigh quotient, with a maximum value in this case evaluated to be simply \cite{rafaely2010book}:

\begin{equation}\label{EQ:dnOPT-Q}
\mathbf{d} = \left[1, 1,..., 1 \right]^T
\end{equation}

In the spherical microphone array literature, this is referred to as regular beam pattern, or plane-wave decomposition, representing directivity functions having a closed-form expression \cite{rafaely2010book}:

\begin{equation}\label{EQ:B-OPT-Q}
B(\Theta) = \frac{N+1}{4\pi(\cos\Theta-1)}[P_{N+1}(\cos\Theta)-P_N(\cos\Theta)]
\end{equation}

also referred to as hyper-cardioid beam pattern, with the maximal directivity factor of $(N+1)^2$.

\subsection{Maximum WNG}\label{SEC:maxWNG}

In a similar manner, the weights $d_n$ that maximize the WNG can also be computed. Equation (\ref{EQ:WNG-dn}) can be written in a matrix form as:

\begin{equation}\label{EQ:WNG-Rayleigh}
 WNG = \frac{\mathbf{d}^H [\mathbf{a}^T \mathbf{a}] \mathbf{d}}{\mathbf{d}^H [\mathrm{diag}(\mathbf{c})] \mathbf{d}}
\end{equation}

with

\begin{equation}\label{EQ:c}
\mathbf{c} = \left[1/b_0(kr_0), 3/b_1(kr_0),..., (2N+1)/b_N(kr_0) \right]^T
\end{equation}

and with a maximum value in this case achieved with weights given by \cite{rafaely2010book}:

\begin{equation}\label{EQ:dnOPT-WNG}
d_n(k) = \frac{4\pi |b_n(kr_0)|^2}{\sum_{n=0}^N |b_n(kr_0)|^2 (2n+1)}
\end{equation}

Again, this is similar to the result obtained for the spherical microphone array.

\subsection{Other beam pattern designs}\label{SEC:DC}

Due to the similarity in the directivity function, directivity factor, and WNG, between the spherical loudspeaker array presented above and the spherical microphone array developed elsewhere, a range of beam pattern design methods can be applied to the spherical loudspeaker array, see, for example \cite{rafaely2010book}. These include, among other, the Dolph-Chebyshev beam pattern, providing optimal trade-off between main-lobe width and side-lobe level, and other optimal design methods.

\section{Experimental study}\label{SEC:experiment}

The aim of this section is to provide an experimental examination of the beamforming design methods presented in this paper. The examination is based upon comparison of measured beam patterns and simulated beam patterns. The simulated beam patterns are generated by using some of the analytical design method presented in this paper to compute beamforming weights and apply them to a computer model of a spherical loudspeaker array, as presented in this paper. The model represents an experimental spherical loudspeaker array system, and so the same weights are applied to the experimental system to produce beam patterns evaluated by microphones measuring the sound pressure away from the spherical loudspeaker array. The measured and simulated beam patterns are then compared.

The experimental system includes a spherical loudspeaker array of radius $r_0=0.15\,$m, with 12 individual loudspeaker units mounted on it's surface, in a dodecahedron arrangement. The loudspeaker array is designed and produced by the Institute of Technical Acoustics, Aachen university, and includes power amplifiers to drive each loudspeaker unit individually. The power amplifiers are connected to a two-channel sound card via a switching circuit, such that each loudspeaker unit can be separately driven by the sound card.

A microphone attached to a rotating system is used to spatially sample the sound pressure radiated by the loudspeaker array. The microphone positions followed the Gaussian sampling scheme \cite{rafaely2005tsap}, with a total of $242$ samples, positioned at a radius of $r=0.57\,$m, achieving a spherical harmonic order of $N=10$ at the analysis sphere.

Once the microphone is positioned in place, the impulse response between each loudspeaker unit and the microphone is measured using a linearly swept-sine signal of a duration of 4 seconds, in the range of $0-1500\,$Hz. A sampling frequency $3000\,$Hz was employed by the measurement system. A sound card connected to a computer running MATLAB was used to play and record the signals. An entire session includes measuring and saving the impulse response data for each microphone position and each loudspeaker unit, giving a total of $242 \times 12 = 2904$ impulse response measurements during a complete session. The entire experiment was performed at the anechoic chamber, acoustics laboratory, Ben-Gurion University of the Negev, having inner dimensions of $2\,$m, certified as anechoic from $300\,$Hz.

At a frequency of $400\,$Hz, the value of $kr_0$ is about $1.1$, and the value of $kr$ is about $4.2$, and so both the spherical loudspeaker array and the measuring spherical microphone array satisfy $kr_0<2$ and $kr<10$, therefore providing spatial over-sampling in both systems. At a frequency of $1000\,$Hz, $kr_0\approx 2.75$ and $kr\approx 10.45$, above which spatial aliasing is expected to be significant in both systems.

The design framework presented in this paper was used for the computation of $d_n$, from which $w_{nm}$ was computed using Eq. (\ref{EQ:wnm-dn}). Then, $w_l$, the weight assigned to each loudspeaker unit was computed from $w_{nm}$, as detailed below, and applied directly to the measured data to compute the measured beam pattern. With the aim of calculating $w_l$, Eq. (\ref{EQ:wnm}) can be written in a matrix form as:

\begin{equation}\label{EQ:wnmMAT}
\mathbf{w_{nm}} = \mathbf{G} \mathbf{Y} \mathbf{w},
\end{equation}

where the $(N+1)^2\times 1$ vector of beamforming weights at wave number $k$ is defined by:

\begin{equation}
\mathbf{w_{nm}} = [w_{00}(k),w_{1(-1)}(k),w_{10},w_{11},...,w_{NN}(k)]^T.
\end{equation}

The spherical harmonics matrix $\mathbf{Y}$ of size $(N+1)^2\times L$ has element at row $q$ and column $l$ given by:

\begin{equation}
Y_{ql} = [Y_n^m(\theta_l,\phi_l)]^*,\quad
q=n^2+n+m,\,\,n=0,...,N,\,\,m=-n,...,n,\,\,l=1,...,L.
\end{equation}

Matrix $\mathbf{G}$ of size $(N+1)^2\times (N+1)^2$ is given by:

\begin{equation}
\mathbf{G} = \mathrm{diag}(g_0,g_1,g_1,g_1,...,g_N)
\end{equation}

and the $L\times 1$ vector $\mathbf{w}$ is given by

\begin{equation}
\mathbf{w} = [w_1(k),w_2(k),...,w_L(k)]^T
\end{equation}

Having designed $d_n$ to achieve a desired beam pattern, Eq. (\ref{EQ:wnm-dn}) is used to derive $w_{nm}$ given the look direction $(\theta_0,\phi_0)$, from which the weights assigned to each loudspeaker unit, $w_l$, is computed by:

\begin{equation}\label{EQ:wMAT}
\mathbf{w} = \mathbf{Y}^\dagger \mathbf{G}^{-1} \mathbf{w_{nm}}
\end{equation}

where $\mathbf{Y}^\dagger$ is the pseudo-inverse of $\mathbf{Y}$.

\begin{figure}[ht]
\centering
\includegraphics[scale=0.4]{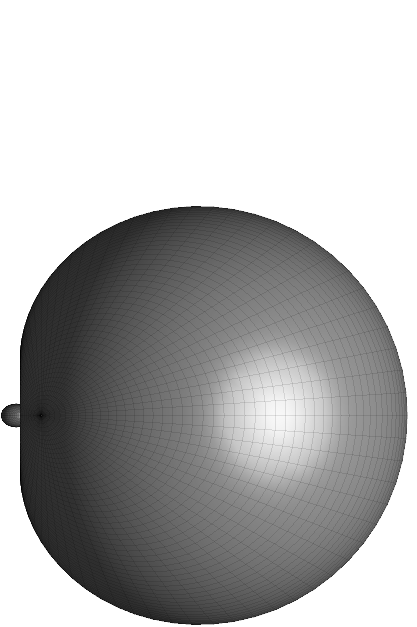}
\includegraphics[scale=0.4]{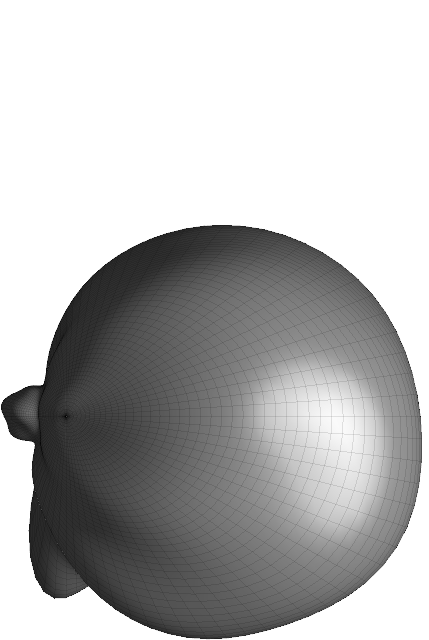}
\caption{Balloon plot of the directivity function, designed using the maximum WNG method with $N=2$, for an operating frequency of $400\,$Hz. Left: simulated, right: measured.}
\label{FIG:baloon400}
\end{figure}

\begin{figure}[ht]
\centering
\includegraphics[scale=0.4]{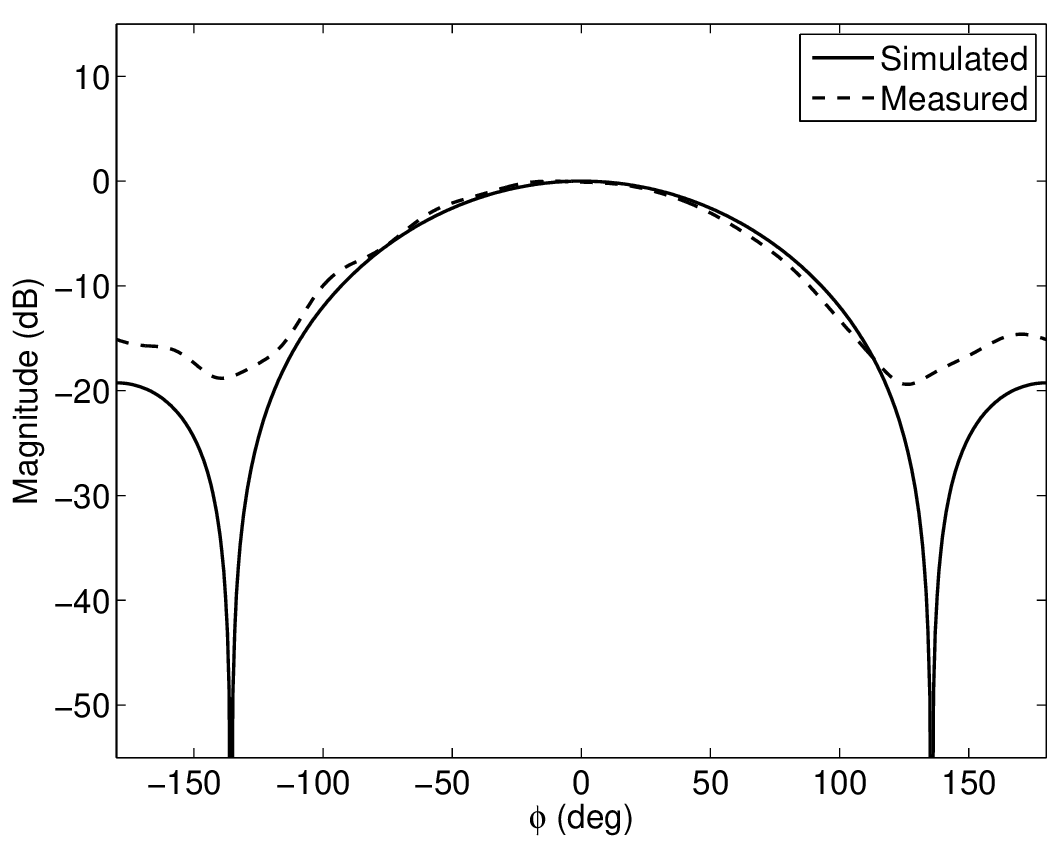}
\caption{Same as in Fig. \ref{FIG:baloon400}, but showing a cross-section along $(\pi/2,\phi)$.}
\label{FIG:WP400}
\end{figure}

Beamforming weights $d_n$ were designed as described above, based on the maximum WNG method, as in Eq. (\ref{EQ:dnOPT-WNG}), and the maximum directivity method, as in Eq. (\ref{EQ:dnOPT-Q}). It should be noted that these are only example designs, and other methods as discussed in this paper could also be used. Figure \ref{FIG:baloon400} shows balloon plots of the simulated and measured beam patterns for a design frequency of $400\,$Hz, using the maximum WNG method. Figure \ref{FIG:WP400} shows a cross-section along the azimuth angle $\theta$ for elevation angle $\phi=\pi/2$. Note that because in this paper the design methods produce axis-symmetric beam pattern, any cross-section intersecting the look direction can be presented. The figures show a reasonable similarity between simulated and measured beam patterns, validating the proposed design framework. Improving the agreement between simulated and measured beam patterns may require a more accurate matching between transducers. Also, because the distance of $r=0.57\,$m of the microphones cannot be considered far-field, the designed weights were modified to account for this near-field effect according to Eq. (\ref{EQ:pressure-wnm}).

\begin{figure}[ht]
\centering
\includegraphics[scale=0.4]{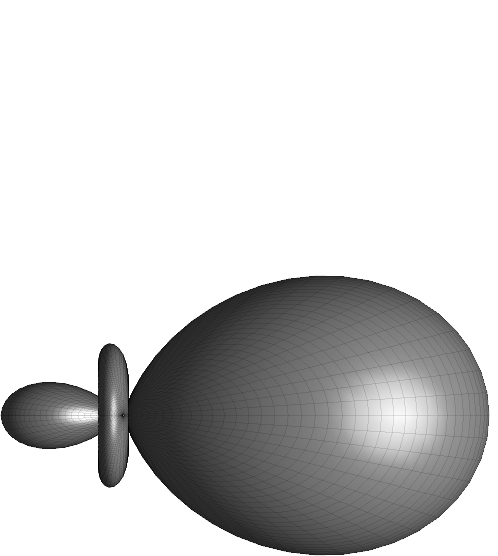}
\includegraphics[scale=0.4]{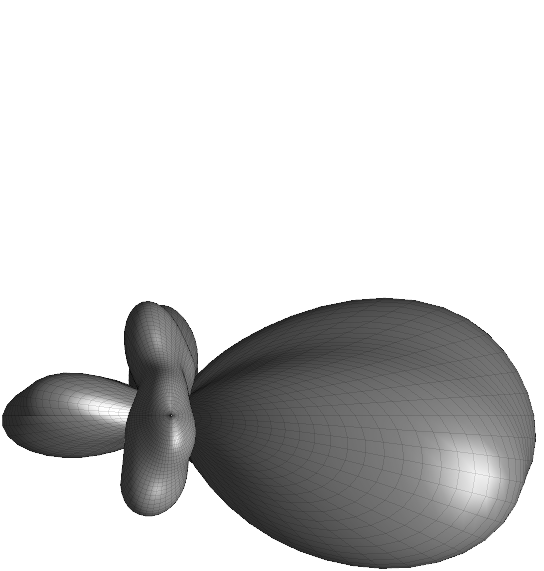}
\caption{Balloon plot of the directivity function, designed using the maximum directivity method with $N=2$, for an operating frequency of $1000\,$Hz. Left: simulated, right: measured.}
\label{FIG:baloon1000}
\end{figure}

\begin{figure}[ht]
\centering
\includegraphics[scale=0.4]{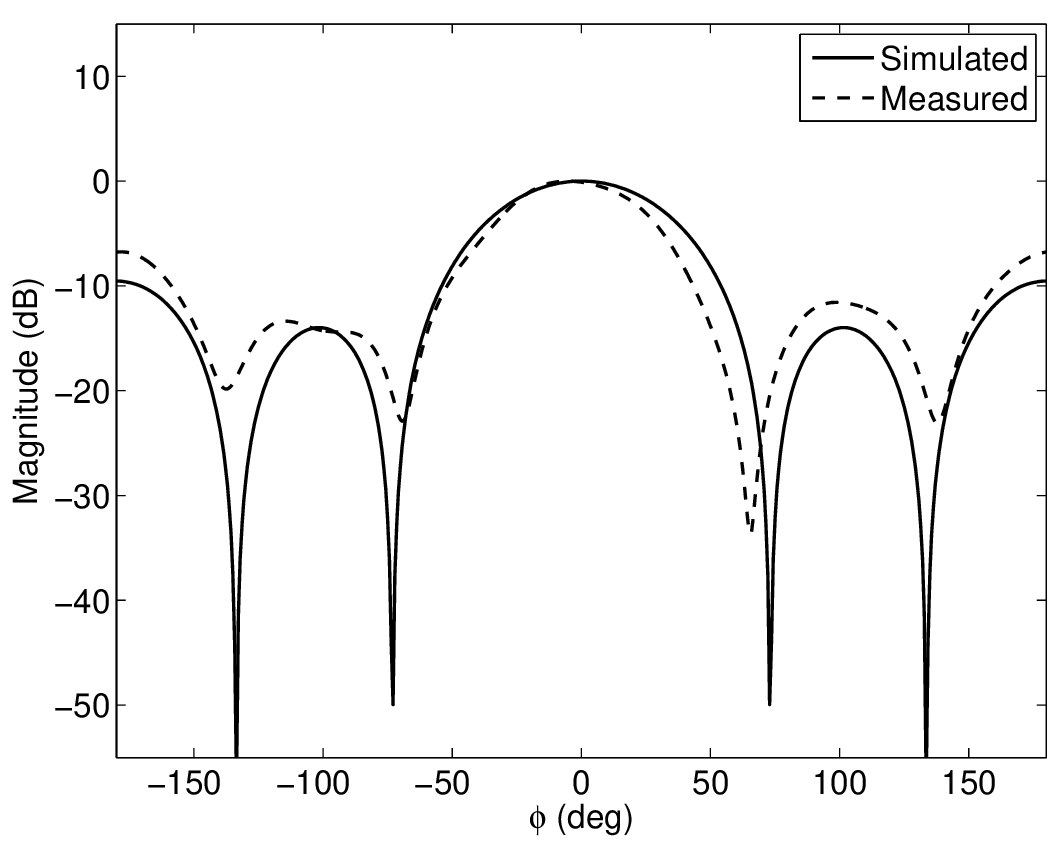}
\caption{Same as in Fig. \ref{FIG:baloon1000}, but showing a cross-section along $(\pi/2,\phi)$.}
\label{FIG:WP1000}
\end{figure}

Figures \ref{FIG:baloon1000} and \ref{FIG:WP1000} show similar results for $1000\,$Hz, using the maximum directivity design method, with a different beam pattern, which has a narrower main lobe. The measured beam pattern is similar to the simulated one, once again validating the design framework.

\section{Conclusion}

This paper presented an efficient beamforming framework for spherical loudspeaker arrays, facilitating optimal, closed-form beam pattern design, with independent steering. The paper derives beamforming equations for the spherical loudspeaker array, showing similarity to spherical microphone arrays configured around a rigid sphere. This similarity facilitates the use of a wide range of beamforming methods already developed for spherical microphone arrays. The design framework is then employed for beamforming with an experimental spherical loudspeaker array system, validating the theoretical results. The proposed framework can be used to produce directional radiation patterns with the spherical loudspeaker array in a wide range of applications.

\section{Acknowledgement}
This work was supported in part by the Ministry of Industry and Trade, grant no. 40161.

\bibliographystyle{IEEEtran}
\bibliography{publications_rafaely_journal,IEEE_LS_beamforming}

\begin{thebibliography}{10}
\providecommand{\url}[1]{#1}
\csname url@samestyle\endcsname
\providecommand{\newblock}{\relax}
\providecommand{\bibinfo}[2]{#2}
\providecommand{\BIBentrySTDinterwordspacing}{\spaceskip=0pt\relax}
\providecommand{\BIBentryALTinterwordstretchfactor}{4}
\providecommand{\BIBentryALTinterwordspacing}{\spaceskip=\fontdimen2\font plus
\BIBentryALTinterwordstretchfactor\fontdimen3\font minus \fontdimen4\font\relax}
\providecommand{\BIBforeignlanguage}[2]{{%
\expandafter\ifx\csname l@#1\endcsname\relax
\typeout{** WARNING: IEEEtran.bst: No hyphenation pattern has been}%
\typeout{** loaded for the language `#1'. Using the pattern for}%
\typeout{** the default language instead.}%
\else
\language=\csname l@#1\endcsname
\fi
#2}}
\providecommand{\BIBdecl}{\relax}
\BIBdecl

\bibitem{warusfel2001}
O.~Warusfel and N.~Misdariis, ``Directivity synthesis with a 3d array of loudspeakers: application for stage performance,'' in \emph{Proceedings of the COSTG-6 Conference on Digital Audio Effects (DAFX-Ol), Limerick, Ireland}, 2001.

\bibitem{avizienis2006}
R.~Avizienis, A.~Freed, P.~Kassakian, and D.~Wessel, ``A compact 120 independent element spherical loudspeaker array with programmable radiation patterns,'' in \emph{Proceedings of the 120th Audio Engineering Society Convention, Paris}, no. 6783, 2006.

\bibitem{rafaely2009jasa}
B.~Rafaely, ``Spherical loudspeaker array for local active control of sound,'' \emph{J. Acoust. Soc. Am.}, vol. 125, no.~5, pp. 3006--3017, May 2009.

\bibitem{zotter2007ICA}
F.~Zotter and R.~H\"{o}ldrich, ``Modeling of radiation synthesis with spherical loudspeaker arrays,'' in \emph{Proceedings of the 19th International Congress on Acoustics, Madrid}, no. COM-01-006, September 2007.

\bibitem{williams1999}
E.~G. Williams, \emph{Fourier acoustics: sound radiation and nearfield acoustical holography}.\hskip 1em plus 0.5em minus 0.4em\relax New York: Academic Press, 1999.

\bibitem{pasqual2010}
A.~M. Pasqual, J.~R. de~Franca~Arruda, and P.~Herzog, ``{Application of Acoustic Radiation Modes in the Directivity Control by a Spherical Loudspeaker Array},'' \emph{Acta Acustica united with Acustica}, vol.~96, no.~1, pp. 32--42, 2010.

\bibitem{pollow2009}
M.~Pollow and G.~K. Behler, ``Variable directivity for platonic sound sources based on spherical harmonics optimization,'' \emph{Acta Acoustics united with Acoustica}, vol.~95, pp. 1082--1092, 2009.

\bibitem{meyer2002}
J.~Meyer and G.~W. Elko, ``A highly scalable spherical microphone array based on an orthonormal decomposition of the soundfield,'' \emph{Proceedings ICASSP 2002}, vol.~II, pp. 1781--1784, 2002.

\bibitem{rafaely2005tsap}
B.~Rafaely, ``Analysis and design of spherical microphone arrays,'' \emph{IEEE Trans. Speech Audio Proc.}, vol.~13, no.~1, pp. 135--143, January 2005.

\bibitem{rafaely2010book}
B.~Rafaely, Y.~Peled, M.~Agmon, D.~Khaykin, and E.~Fisher, ``Spherical microphone array beamforming,'' in \emph{Speech Processing in Modern Communications: challenges and perspectives}, I.~Cohen, J.~Benesty, and S.~Gannot, Eds.\hskip 1em plus 0.5em minus 0.4em\relax Berlin: Springer-Verlag, 2010, ch.~11, pp. 281--305.

\bibitem{zotter2007DAGA}
F.~Zotter and R.~H\"{o}ldrich, ``Modeling a spherical loudspeaker system as a multipole source,'' in \emph{Proceedings of the 33rd German Annual Conference on Acoustics, Stuttgart}, March 2007.

\bibitem{arfken2001}
G.~Arfken and H.~J. Weber, \emph{Mathematical methods for physicists}, 5th~ed.\hskip 1em plus 0.5em minus 0.4em\relax San Diego: Academic Press, 2001.

\bibitem{kinsler1999}
L.~E. Kinsler, A.~R. Frey, A.~B. Coppens, and J.~V. Sanders, \emph{Fundamentals of Acoustics}, 4th~ed.\hskip 1em plus 0.5em minus 0.4em\relax New York: John Wiley \& Sons, 1999.

\bibitem{vantrees2002}
H.~L. Van-Trees, \emph{Optimum Array Processing (Detection, Estimation, and Modulation Theory, Part IV)}, 1st~ed.\hskip 1em plus 0.5em minus 0.4em\relax Wiley-Interscience, 2002.

\bibitem{rafaely2005spl}
B.~Rafaely, ``Phase-mode versus delay-and-sum spherical microphone array processing,'' \emph{IEEE Sig. Proc. Let.}, vol.~12, no.~10, pp. 713--716, October 2005.

\end{thebibliography}



\end{document}